%% file: ms.tex
\documentclass[11pt]{article}\usepackage[]{graphicx}\usepackage[]{xcolor}
\makeatletter
\def\maxwidth{ %
  \ifdim\Gin@nat@width>\linewidth
    \linewidth
  \else
    \Gin@nat@width
  \fi
}
\makeatother

\definecolor{fgcolor}{rgb}{0.345, 0.345, 0.345}

\usepackage{framed}
\makeatletter
 {\par\unskip\endMakeFramed%
 \at@end@of@kframe}
\makeatother

\definecolor{shadecolor}{rgb}{.97, .97, .97}
\definecolor{messagecolor}{rgb}{0, 0, 0}
\definecolor{warningcolor}{rgb}{1, 0, 1}
\definecolor{errorcolor}{rgb}{1, 0, 0}
\newenvironment{knitrout}{}{} 

\usepackage{alltt}
\usepackage{fullpage}

\usepackage[sectionbib]{natbib}

 \newcommand\statSinica[1]{}
 \newcommand\arxiv[1]{#1}

\input{header.tex}

\usepackage{amsmath,amssymb}
\usepackage{graphicx}
\usepackage{enumerate}
\usepackage{natbib}
\usepackage{soul}
\usepackage{xcolor}

\usepackage{url} 

\IfFileExists{upquote.sty}{\usepackage{upquote}}{}
\begin{document}


\statSinica{\statSinicaSetup}

\centerline{\statSinica{\large}\arxiv{\Large}\bf An iterated block particle filter for inference on coupled dynamic}

\arxiv{\vspace{1mm}}

\centerline{\statSinica{\large}\arxiv{\Large}\bf systems with shared and unit-specific parameters}

\arxiv{\vspace{2mm}}

\centerline{Edward L. Ionides$^1$, Ning Ning$^2$ and Jesse Wheeler$^1$}

\arxiv{\vspace{1mm}}

{\small
\centerline{\small University of Michigan, Department of Statistics$^1$}
\centerline{\small Texas A\&M University, Department of Statistics$^2$}
}
 \vspace{.25cm} \fontsize{9}{11.5pt plus.8pt minus.6pt}\selectfont

\statSinica{\statSinicaSubmission}


\begin{quotation}
\noindent {\it Abstract:}
We consider inference for a collection of partially observed, stochastic, interacting, nonlinear dynamic processes.
Each process is identified with a label called its unit, and our primary motivation arises in biological metapopulation systems where a unit corresponds to a spatially distinct sub-population.
Metapopulation systems are characterized by strong dependence through time within a single unit and relatively weak interactions between units, and these properties make block particle filters an effective tool for simulation-based likelihood evaluation.
Iterated filtering algorithms can facilitate likelihood maximization for simulation-based filters.
We introduce an iterated block particle filter applicable when parameters are unit-specific or shared between units.
We demonstrate this algorithm by performing inference on a coupled epidemiological model describing spatiotemporal measles case report data for twenty towns.

\statSinica{\statSinicaKeywords}

\end{quotation}\par

\def\thefigure{\arabic{figure}}
\def\thetable{\arabic{table}}

\renewcommand{\theequation}{\thesection.\arabic{equation}}

\statSinica{
  \fontsize{12}{14pt plus.8pt minus .6pt}\selectfont
}

\date{}

\section{Introduction}
\label{sec:intro}

Statistical inference for high-dimensional partially-observed nonlinear dynamic systems arises in various scientific contexts.
Massive models and datasets are considered in the geophysical sciences, carried out under the name of data assimilation \citep{evensen09book}.
Population models in ecology and epidemiology can be characterized by high levels of stochasticity, nonlinearity, measurement error, and model uncertainty, leading to challenges of a somewhat different nature than geophysical models.
In addition, biological population systems may sometimes have a low population count due to a local introduction or fade-out of one or more constituent species.
Low population situations may require consideration of integer-valued counts rather than continuous population approximations.
Collections of biological populations measured at different spatial locations may have spatial interactions in addition to local population dynamics; such collections are called a metapopulation.
The study of spatiotemporal disease dynamics has motivated research into inference for metapopulation systems \citep{xia04,li20,park20,ionides21,cauchemez08,bjornstad19}.

Statistical inference for partially observed nonlinear biological systems was, until recently, an open methodological challenge even in the time series case \citep{bjornstad01}.
Advances in Monte Carlo methods based on the particle filter have made inference accessible on many low-dimensional problems \citep{doucet11,kantas15,king16} but the curse of dimensionality \citep{bengtsson08} prevents applicability of the basic particle filter on metapopulations with more than a few units.
Methods based on improving the proposal distribution for the particle filter may not fully resolve the curse of dimensionality \citep{snyder15}.
Situations where previous Monte Carlo methods can provably beat the curse of dimensionality may have limited applicability \citep{beskos17,park20,ionides21}.
Consequently, state-of-the-art scientific analysis for metapopulation dynamics has depended on ensemble Kalman filter (EnKF) methods \citep{li20}.
EnKF algorithms scale well but are founded on an approximation that can be unsuitable for discrete populations with fade-out and re-introduction dynamics and other highly non-Gaussian features \citep{ionides21}.

In Sec.~\ref{sec:alg}, we propose an algorithm for inference on metapopulation dynamics which we call an iterated block particle filter (IBPF).
IBPF combines the iterated filtering likelihood maximization technique of \citet{ionides15} with the block particle filter (BFP) of \citet{rebeschini15}.
Iterated filtering algorithms use parameter perturbations to coerce a filtering algorithm into exploring the parameter space.
BPF addresses the curse of dimensionality by modifying the resampling step of a particle filter to resample independently on blocks that form a partition of the collection of units.
A previous IBPF algorithm was developed by \citet{ning21-ibpf} for the particular case that all estimated parameters are unit-specific, which is to say that the dynamics and measurement process for a unit $u$ are determined by a vector of parameters $\psi_u$ specific to unit $u$.
The formal meaning of this assertion is given, together with the pseudocode for our algorithm, in Sec.~\ref{sec:alg}.
We propose an extension of their IBPF which additionally permits the estimation of a vector of shared parameters, $\phi$.
In this case, the full parameter vector is $\theta=(\phi,\psi_{1:U})$ where the $U$ units are named $\{1,\dots,\Unit\}$ which we denote by $\seq{1}{\Unit}$.
\citet{ning21-ibpf} developed theoretical justification for their algorithm, but our extension of IBPF currently carries only empirical support.

There may be scientific interest concerning which parameters in metapopulation systems are best understood as unit-specific, and which can reasonably be modeled as shared between units.
Another relevant possibility is that a parameter may differ between units as a shared function of unit-specific covariates, which formally is a special case of a shared parameter.
Addressing these issues is also a prerequisite for studying questions about the coupling of metapopulation systems via model-based inference from spatiotemporal data.
We will show empirically, in Sec.~\ref{sec:data}, that our {\ibpf} algorithm is applicable to an inference challenge in epidemiological metapopulation dynamics.
Our demonstration focuses on a dataset for weekly measles incidence in 20 towns in the United Kingdom (UK) during the pre-vaccination era \citep{he10} modeled using a previously studied metapopulation model \citep{park20,ionides21}.
Measles case reports have been a longstanding benchmark problem for inference on biological dynamics, motivating the development of time series methodology and, more recently, the progression from single populations to metapopulation systems.
Unlike previous attempts on sequential Monte Carlo inference for metapopulation models, we show that our algorithm can provide practical plug-and-play, likelihood-based inference when parameters are either shared between units or differ between units.
Demonstrating a solution to this open problem provides numerical evidence substituting for numerical comparisons against alternative methods.

Our data analysis results do not fully resolve open questions about what models for coupling between towns are supported by the data, and which parameters should be modeled as unit-specific.
Rather, we demonstrate steps toward this research goal.
We use a simulation study, in Sec.~\ref{sec:sim}, to show that our methodology can deliver a good approximation to the maximum likelihood estimate when fitting the model used to simulate the data.
This reassurance allows us to interpret our data analysis results as evidence of model misspecification, providing a guide for future investigations of these data as well as tools to carry out those investigations.

Optimization of high-dimensional, non-convex and potentially multi-modal functions, evaluated using stochastic methods, is not straightforward even when it is possible to evaluate the function within an acceptable level of error.
Therefore, we discuss approaches that assist the noisy likelihood searches, and suggest diagnostic plots to assess their success.

\section{The {\ibpf} algorithm for likelihood maximization}
\label{sec:alg}

A latent Markov process is denoted by $\{\myvec{X}_{\time},\time\in \seq{0}{\Time}\}$, with $\myvec{X}_{\time}=X_{1:\Unit,\time}$ taking values in a product space $\Xspace^\Unit$.
We define set-valued subscripts by $X_A=\{X_a,a\in A\}$ and $X_{A,B}=\{X_{a,b}, a\in A, b\in B\}$.
The discrete time process $\myvec{X}_{0:\Time}$ may arise from a continuous time Markov process $\{\myvec{X}(t), t_0\le t\le t_{\Time} \}$ observed at times $t_{1:\Time}$, and in this case we set $\myvec{X}_\time=\myvec{X}(t_\time)$.
The initial value $\myvec{X}_{0}$ may be stochastic or deterministic.
Observations are made on each unit, modeled by an observable process $\myvec{Y}_{1:\Time}=Y_{1:\Unit,1:\Time}$ which takes values  at each time $\time$ in a product space $\Yspace^\Unit$.
Observations are modeled as conditionally independent given the latent process.
The conditional independence of measurements applies over both time and the unit structure, so the collection $\big\{Y_{\unit\comma\time},\unit\in\seq{1}{\Unit},\time\in\seq{1}{\Time}\big\}$ is conditionally independent given
$\big\{X_{\unit\comma\time},\unit\in\seq{1}{\Unit},\time\in\seq{1}{\Time}\big\}$.
We suppose the existence of a joint density $f_{\myvec{X}_{0:\Time},\myvec{Y}_{1:\Time}}$  for $X_{1:\Unit,0:\Time}$ and $Y_{1:\Unit,1:\Time}$ with respect to some appropriate measure, following a notational convention that the subscripts of $f$ denote the joint or conditional density under consideration.
We suppose that $f$ depends on a real-valued parameter vector $\theta=(\phi,\psi_{1:U})$, which we write as $\theta=\theta_{1:D}$ when we wish to concatenate the component shared and unit-specific parameters into a single vector of length $D$.
The data are $\data{y}_{\unit\comma\time}$ for unit $\unit$ at time $\time$.
This model is a special case of a  partially observed Markov process (POMP) also known as a state space model or hidden Markov model.
The additional unit structure, not generally required for a POMP, is appropriate for modeling interactions between units characterized by a spatial location, and so we call the model a SpatPOMP.
For metapopulation models, the units are not generally arranged on a spatial grid but comprise a collection of spatially distributed population centers.

A numerical challenge of fundamental statistical relevance is the maximization of the log-likelihood function of the data given the model, $\loglik(\theta)=\log f_{\myvec{Y}_{1:\Time}}(\data{\myvec{y}}_{1:\Time}\giventh \theta)$.
Numerical evaluation of the likelihood function is closely related to the filtering problem of evaluating $f_{\myvec{X}_n|\myvec{Y}_{1:n}}(\myvec{x}_n\given \data{\myvec{y}}_{1:n}\giventh \theta)$.
If the dynamic model is extended to include the parameters as latent variables, the filtering problem leads to the Bayesian posterior distribution, though regularization is required to make the calculation numerically tractable using Monte Carlo methods \citep{kitagawa98,janeliu01}.
Iterating this Bayesian calculation recursively targets a maximum likelihood estimate (MLE), a strategy known as data cloning \citep{lele07,lele10}.
Adding noise to perturb the parameters in the extended model at each time point can stabilize the numerics while retaining the capability to approximate the MLE \citep{ionides06-pnas,ionides11,ionides15}.
Many variations on this idea have been developed using different filter methods \citep{park20,li20,ionides21,manoli15} or employing different perturbation systems \citep{doucet13,nguyen17}.

A direct approach to iterating BPF for parameter estimation is to resample the extended model independently on each block, giving rise to separate collections of parameters for each block.
\citet{ning21-ibpf} proved that this IBPF algorithm targets the MLE for the special case where each parameter is localized to an individual unit, i.e., when all parameters are unit-specific.
Formally, we say that a parameter for a discrete-time SpatPOMP is specific to a unit $u$ if it is involved in specifying the measurement density $f_{Y_{u,n}|X_{u,n}}$ or transition density $f_{X_{u,n+1}|\myvec{X}_n}$ for some $n$, and it is not involved in $f_{Y_{v,m}|X_{v,m}}$ or $f_{X_{v,m+1}|\myvec{X}_m}$ for any $v\neq u$ and any $m$.
For a continuous-time SpatPOMP, we replace the requirement on $f_{X_{u,n+1}|\myvec{X}_n}$ with an equivalent requirement on a numerical solution over a small time increment $\delta$, as in equation \eqref{eq:extension} below.
A parameter which is not unit-specific is said to be shared between units.
There are intermediate possibilities, where a parameter is shared for only a subset of all units, but such situations are not explored further here.
The special case where all parameters are unit-specific may occur sometimes, but models typically have some shared parameters which arise in transition densities and/or measurement densities for multiple units.

Our approach to iterated filtering for shared parameters involves constructing an extended model within which the shared parameters is represented as a unit-specific parameter that happens to be constant across units.
We construct a spatiotemporally extended model by supposing that a numerical solution for the transition from time $t$ to time $t+\delta$ for each unit $u$ has a functional form
\begin{equation}
\label{eq:extension}
  X_{\unit}(t+\delta) = X_\unit(t) + Q_u
    \big(\myvec{X}(t),\myvec{\eta}_t,\phi,\psi_{\unit},t,\delta \big),
\end{equation}
where the random vector $\myvec{\eta}_t=\eta_{1:\Unit,t}$ is shared for all $u\in \seq{1}{U}$ and does not depend on $\theta$.
If a representation \eqref{eq:extension} exists, an extended model is defined by replacing $\phi$ with $\phi_{\unit}(t)$ and $\psi_u$ with $\psi_{u}(t)$.
Equation \eqref{eq:extension} implicitly defines a continuous-time extended model by the limit as $\delta\to 0$, when that limit exists, but for simulation-based methods the numerical solution is of more immediate concern than this limit.
Admitting a minor abuse of notation, we subsequently use a density $f$ to denote both the original model and its extension for spatiotemporally varying parameters, with the context determining which one is intended.

In some situations, the extended model could have problematic properties.
For example, it could break conservation laws obeyed by the original system.
In other situations, the extended model may make scientific sense in its own right.
For example, in biological metapopulation systems it might be scientifically meaningful to consider a model where there is variation over space and time in the parameters describing the local dynamics.
Here, we are focusing on the hypothesis that there are some parameters that are fixed across space and time, but the specification in \eqref{eq:extension} requires this hypothesis to be nested within a more flexible alternative.


\begin{algorithm}[H]
\setstretch{1.35}
  \caption{
  \textbf{IBPF}. \newline
  {\bf Inputs}:
  simulator for the extended model, $f_{\myvec{X}_{\time}|\myvec{X}_{\time-1}}(\myvec{x}_{\time}\given \myvec{x}_{\time-1}\giventh\myvec{\theta})$ and initialization, $f_{\myvec{X}_0}(\myvec{x}_0\giventh\myvec{\theta})$;
  evaluator for $f_{{Y}_{\unit,\time}|{X}_{\unit,\time}}({y}_{\unit,\time}\given {x}_{\unit,\time}\giventh\theta)$;
  data, $\data{\myvec{y}}_{1:\Time}$;
  number of particles, $\Np$;
  blocks, $\blocklist_{1:\Block}$;
  initial parameter swarm with decomposition into shared and unit-specific parameters, $\Theta^{0,\np}_{\unit}=(\Phi^{0,\np}_{\unit},\Psi^{0,\np}_{\unit})$;
  random walk perturbation, $\sigma_{d,\time}$; cooling rate, $a$; number of iterations, $M$; spatial autoregression, $\spatReg$.
}\label{alg:ibpfilter}
\For{$\nit\ \mathrm{in}\ \seq{1}{\Nit}$}{
  Perturb parameters:
    $\Theta^{F,\nit,\np}_{\unit,0}\sim \normal
    \left(
      \Theta^{\nit-1,\np}_{\unit}\param \sigma^2_{0} a^{2\nit/50}
    \right)$
  \;
  Initialization:  simulate $\myvec{X}_{0}^{F,\np}\sim {f}_{\myvec{X}_{0}}
    \left(\mydot\giventh{\Theta^{F,\nit,\np}_{1:\Unit,0}}\right)$\;
    \For{$\time\ \mathrm{in}\ \seq{1}{\Time}$}{ 
      Perturb parameters:
        $\Theta^{P,\nit,\np}_{\unit,\time}\sim \normal\left(
       \Theta^{F,\nit,\np}_{\unit,\time-1},\sigma^2_\time a^{2\nit/50} \right)$
     \;
      Prediction simulation: $\myvec{X}_{\time}^{P,\np}
        \sim {f}_{\myvec{X}_{\time}|\myvec{X}_{\time-1}}\big(
        \mydot|\myvec{X}_{\time-1}^{F,\np};{\Theta^{P,\nit,\np}_{1:\Unit,\time}}\big)$
      \;
    \For{$\block\ \mathrm{in}\ \seq{1}{\Block}$}{
      Block prediction weights:
            $\displaystyle \blockweight^P_{\time,\np, \block}=
              \prod_{\unit \in \blocklist_{\block}}
              f_{Y_{\unit,\time}|X_{\unit,\time}}
              \big(
              \data{y}_{\unit,\time}\given X^{P,\np}_{\unit,\time}
                \giventh \Theta^{P,\nit,\np}_{\unit,\time} \big)$
            \nllabel{alg:bpfilter:blockweights}
      \;
      Normalize weights:
        $\tilde{\blockweight}_{\time,\np, \block}= \blockweight^P_{\time,\np,\block}\Big/
        \sum_{i=1}^{\Np}\blockweight^P_{\time,i,\block}$\;
      Select resample indices:
        $\resampleIndex_{1:\Np,\block}$ with
        $\prob\left[\resampleIndex_{\np,\block}=\altNp\right] =
        \tilde{\blockweight}_{\time,\np,\block}$
      \;
         $X_{\blocklist_{\block},\time}^{F,\np}=X_{\blocklist_{\block},\time}^{P,\resampleIndex_{\np,\block}}$,
	$\Theta_{\blocklist_{\block},\time}^{F,\np}
	  =\Theta_{\blocklist_{\block},\time}^{P,\resampleIndex_{\np,\block}}
	  =\big(\Phi_{\blocklist_{\block},\time}^{P,\resampleIndex_{\np,\block}},
	    \Psi_{\blocklist_{\block},\time}^{P,\resampleIndex_{\np,\block}}\big)$
         \nllabel{alg:bpfilter:resample}
      \;
      block mean of shared parameters: $\mu_{\block,\time}^{} = \Np^{-1}\sum_{\np=1}^{\Np} \Phi^{F,\np}_{\blocklist_{\block},\time}$
    } 
    overall mean of shared parameters: $\mu_{\time}^{} = \Block^{-1}\sum_{\block=1}^{\Block} \mu_{\block,\time}^{}$
    \;
    autoregressive correction:
    $\Phi^{F,\np}_{\blocklist_{\block},\time} = \Phi^{F,\np}_{\blocklist_{\block},\time} + \spatReg\big( \mu_{\time}^{} - \mu_{\block,\time}^{}  \big)$
  \;
  }
  $\Theta_{\unit}^{\nit,\np}=\Theta^{F,\nit,\np}_{\unit,\Time}$
  \;
} 
\KwOut{
    IBPF parameter swarm, $\Theta_{\unit}^{\Nit,\np}$
}
\end{algorithm}

The IBPF algorithm described in Alg.~\ref{alg:ibpfilter} carries out a block particle filter on this extended model.
The extended parameters are given independent perturbations, but a spatial autoregressive step pulls the values of shared parameters toward their mean over the units.
Indeed, this autoregressive step is the only difference between  Alg.~\ref{alg:ibpfilter} and the algorithm for unit-specific parameters proposed by \citet{ning21-ibpf}.
This algorithm, in turn, is essentially the IF2 algorithm of \citet{ionides15} with the particle filter replaced by BPF.

Alg.~\ref{alg:ibpfilter} assumes implicit loops over $\np\ \text{in}\ \seq{1}{\Np}$ and $\unit\ \text{in}\ \seq{1}{\Unit}$,
$\normal(\mu,\Sigma)$ denotes the normal distribution with mean $\mu$ and variance matrix $\Sigma$, and
$\sigma_\time$ is a $D\times D$ diagonal matrix with entries $\sigma_{d,\time}$.
The blocks, $\blocklist_{1:\Block}$, are a partition of $\seq{1}{\Unit}$.

The pseudocode in Alg.~\ref{alg:ibpfilter} represents our implementation of {\ibpf} as the R function \code{ibpf} which we have contributed to the open-source package \code{spatPomp} \citep{asfaw21github,asfaw21arxiv}.
Additionally, the source code for all the results in this article is available at \url{https://github.com/ionides/ibpf_article}.
Various generalizations of this implementation are possible.
For example, iterated filtering theory does not rely on parameter perturbations following the normal distribution \citep{ionides15} though in practice we transform parameters to facilitate this convenient choice (Sec.~\supSecAlg).

\subsection{Algorithmic parameters}

Model parameters are optimized on a transformed scale for which unit variation is scientifically meaningful.
In practice, this means working with positive parameters on a log scale, and $(0,1)$ interval-valued parameters on a logistic scale.
We follow standard iterated filtering practice by using independent random walks for each parameter on this transformed scale \citep{king16}.
We discovered that the large number of parameters following a random walk, in the presence of unit-specific parameters, can require considerably smaller random walk standard deviations than the values around $\sigma_{d,n}=0.02$ (i.e., 2\% perturbation per time point) that have been employed for iterated filtering of time series models.
After experimentation, we used $\sigma_{d,n}=0.005$ for the initial search and $\sigma_{d,n}=0.00125$ for subsequent refinements.
Occasionally, a parameter which can be estimated precisely from the data can benefit substantially from a smaller perturbation.
This is the case for one parameter in our measles analysis, an exponent $\alpha$ for which the scale of uncertainty is an order of magnitude smaller than other parameters; therefore, $\sigma_{d,n}$ for this parameter was scaled accordingly.
In principle, $\sigma_{d,n}$ can be a function of $n$ as well as the parameter, $d$.
The most common reason for using this flexibility is to avoid perturbing parameters during time intervals when there is no information about them: following an evolutionary analogy, evolution cannot operate effectively when there is mutation but no selection.
For an initial value parameter, meaning one which specifies only the latent process value at the initial time $t_0$, we use $\sigma_{d,n}=0$ for $n\ge 1$ and we doubled the value of $\sigma_{d,n}$ for $n=0$.

We used $J=4000$ particles, and we set the cooling rate parameter to be $a=0.5$, corresponding to a 1\% reduction in the random walk standard deviation at each iteration.
We set $M=100$ optimizations iterations, chosen as an empirically assessed compromise between effort spent on each search and the number of searches conducted.
Additional discussion of algorithmic parameters is given in Sec.~\supSecAlg.
The only additional algorithmic parameter over previous iterated filtering algorithms is the spatial autoregressive parameter, $\spatReg$.
Numerical experimentation suggests that the performance is not sensitive to the choice of $\spatReg>0$ (Supplementary Fig.~\supFigSpatReg).
Subsequently, we used $\spatReg=0.1$.

Applying IBPF to real data forces us to address issues of model development, model misspecification, and performance in the presence of outliers.
Before tackling these issues, we use simulated data to demonstrate the capabilities of the methodology on a correctly specified model.

\begin{knitrout}
\definecolor{shadecolor}{rgb}{1, 1, 1}\color{fgcolor}\begin{figure}

\includegraphics[width=5.5in]{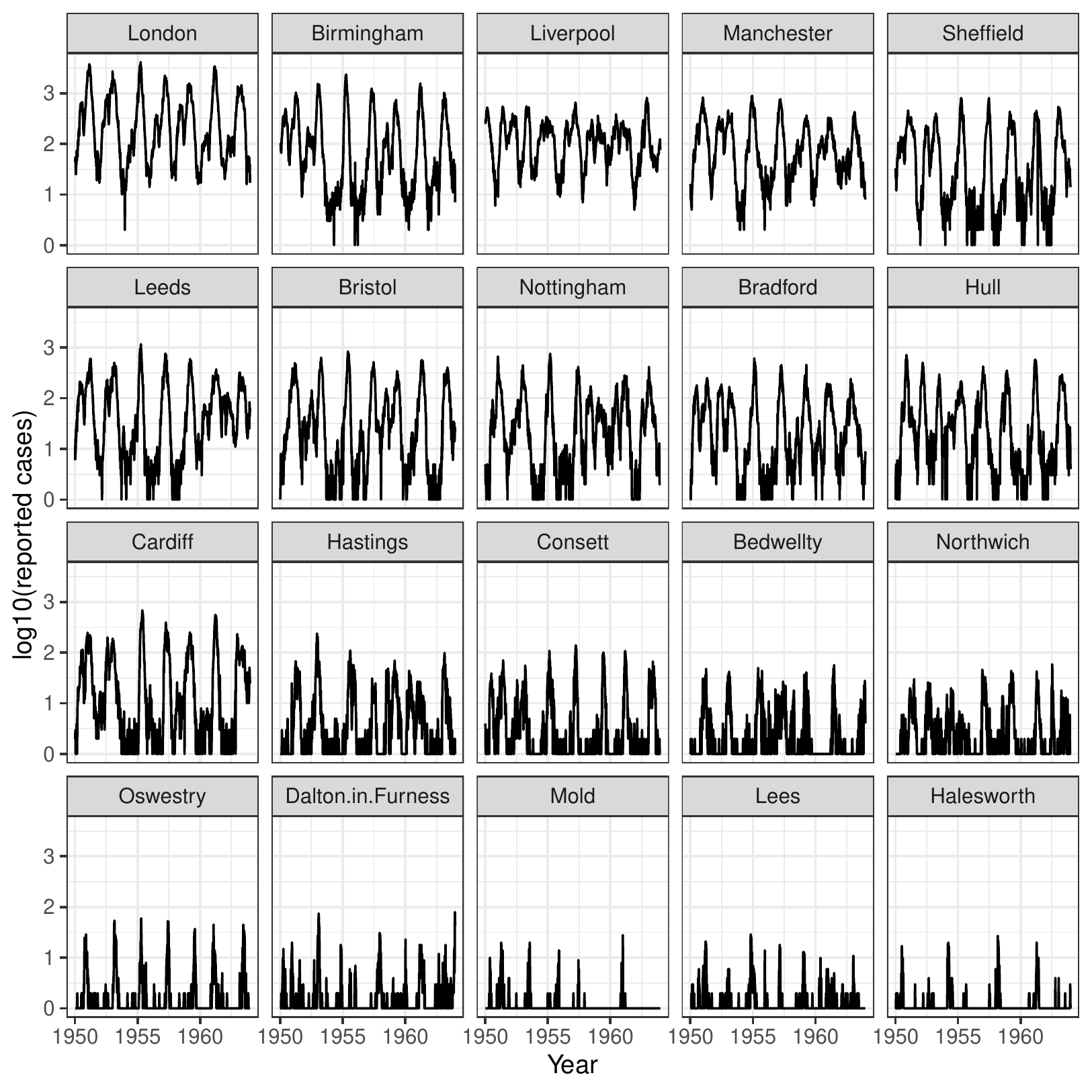} \hfill{}

\caption[Weekly measles case reports for 20 UK towns]{Weekly measles case reports for 20 UK towns.}\label{fig:data-plot}
\end{figure}

\end{knitrout}

\section{Testing {\ibpf} on a measles transmission model}
\label{sec:measles}

Measles transmission has provided a useful example of epidemiological dynamics (and therefore also ecological dynamics, for a host-pathogen ecosystem), with plentiful case report data and relatively simple biology.
Model-based analysis of measles time series data led to advances in understanding of the seasonality of infectious disease \citep{fine82}, critical community size \citep{bartlett57}, the recognition that relatively simple mechanistic models can provide a remarkably good description of the dynamics \citep{earn00}, and much other foundational research on disease dynamics.
Some progress has been on building and fitting spatiotemporal models for measles; e.g., \citet{xia04,eggo11,bjornstad19,becker20}.
However, the lack of suitable methodology to fit and assess a flexible class of coupled models is an obstacle to further progress \citep{becker20}.
Previous methodological work has used spatiotemporal measles models as a test problem \citep{park20,ionides21}, but these methods have fallen short as tools for data analysis due to numerical considerations.
Bearing all this in mind, measles provides a natural testing ground for our new methodologies.
We demonstrate that we now have the tools to carry out likelihood maximization (and therefore, in principle, profile likelihood confidence interval construction and likelihood-based model selection) on mechanistic statistical models that are appropriate for spatiotemporal metapopulation disease data.

\subsection{The measles data, a model and three submodels}
\label{sec:model}

We set ourselves the task of fitting a spatiotemporal model to the case reports for 20 towns studied by \citet{he10}.
We require the ability to handle discrete case counts which vary from zero to thousands of cases per week.
We consider a model with the same structure as \citet{he10}, namely a Markov chain with Gamma noise on the infection rate, but with the addition of a term for transmission between cities.
This requirement limits us to plug-and-play methodologies, which are those that require simulation from the latent process model but not the ability to  evaluate transition densities \citep{breto09,he10}.

Some previous analyses have used counts aggregated over two-week intervals \citep{park20,ionides21} since these were available for more cities, but our goal here is to extend the analysis of \citet{he10}.
Apart from this, our model matches \citet{ionides21} and for completeness we repeat the description here.
We compartmentalize the population of each town into susceptible ($S$), exposed ($E$), infectious ($I$), and recovered/removed ($R$) categories.
The numbers of individuals in each compartment for town $\unit$ at time $t$ are denoted by integer-valued random variables $S_\unit(t)$, $E_\unit(t)$, $I_\unit(t)$, and $R_\unit(t)$.
The population dynamics are written in terms of counting processes $N_{Q_1Q_2,\unit}(t)$ enumerating cumulative transitions in town $\unit$, up to time $t$, from compartment $Q_1$ to $Q_2$.
Here, $Q_1,Q_2\in \{S,E,I,R,B,D\}$ with $B$ denoting a source compartment for immigration or birth, and $D$ a sink compartment for emigration or death.
We enumerate the $U=20$ towns studied by \citet{he10} in decreasing size, so that $\unit=1$ corresponds to London.
Our model is described by the following system of stochastic differential equations, for $\unit=1,\dots, \Unit$,
\begin{equation}
\nonumber
\begin{array}{lllllll}
\displaystyle dS_\unit(t) &=& dN_{BS,\unit}(t) &-& dN_{SE,\unit}(t) &-& dN_{SD,\unit}(t), \\
\displaystyle dE_\unit(t) &=& dN_{SE,\unit}(t) &-& dN_{EI,\unit}(t) &-& dN_{ED,\unit}(t), \\
\displaystyle dI_\unit(t) &=& dN_{EI,\unit}(t) &-& dN_{IR,\unit}(t) &-& dN_{ID,\unit}(t).
\end{array}
\end{equation}
The total population $P_\unit(t)=S_\unit(t)+E_\unit(t)+I_\unit(t)+R_\unit(t)$ is calculated by smoothing census data and is treated as known.
The number of recovered individuals $R_\unit(t)$ in town $\unit$ is therefore defined implicitly.
$N_{SE,\unit}(t)$ is modeled as a negative binomial death process \citep{breto09,breto11} defined by a rate, $\mu_{SE,\unit}(t)$, specified as
\begin{equation}
\nonumber
\mu_{SE,\unit}=
\beta_{\unit}
\left[
  \left( \frac{I_\unit+\iota_u}{P_\unit} \right)^{\alpha_{\unit}}
\hspace{-1mm}
 + \sum_{\altUnit \neq \unit} \frac{v_{\unit\altUnit}}{P_\unit}
  \left\{
    \left(
      \frac{ I_{\altUnit} }{ P_{\altUnit} }
    \right)^{\alpha_{\unit}} -
    \hspace{-1mm}
    \left(
      \frac{I_\unit}{P_\unit}
    \right)^{\alpha_{\unit}}
  \right\}
\right] \frac{d\Gamma_{SE,\unit}}{dt}.
\end{equation}
Here, the time dependence of $\mu_{SE,u}$, $\beta_u$, $I_u$ and $P_u$ is suppressed;
$\alpha_{\unit}$ is a mixing exponent modeling inhomogeneous contact rates within a town;
$\iota_{\unit}$ models immigration of infected individuals;
${d\Gamma_{SE,\unit}}/{dt}$ is gamma white noise, with intensity parameter $\sigma_{SE.\unit}$;
$\beta_{\unit}$, models the seasonal transmission driven by high contact rates between children at school,
\begin{equation}
\nonumber
  \beta_{\unit}(t)=\Bigg\{
  \begin{array}{ll}
\big(1+\amplitude_{\unit}(1-\schoolTermFrac)\schoolTermFrac^{-1} \big)\, \meanBeta_{\unit} & \mbox{ during school term},\\
\big( 1-\amplitude_\unit\big) \, \meanBeta_{\unit} & \mbox{ during vacation},
  \end{array}
\label{eq:term}
\end{equation}
where $\schoolTermFrac = 0.759$ is the proportion of the year taken up by the school terms, $\meanBeta_\unit$ is the mean transmission rate, and $\amplitude_{\unit}$ measures the reduction of transmission during school holidays.

The number of travelers from town $\unit$ to $\altUnit$ is denoted by $v_{\unit\altUnit}$.
Here, $v_{\unit\altUnit}$ is constructed using the gravity movement model of \cite{xia04} given by
\[
v_{\unit\altUnit} = \gravity_{\unit} \cdot \frac{\hspace{2mm} \overline{\dist}\hspace{2mm}}{\hspace{1.5mm}\overline{\pop}^2\hspace{1mm}} \cdot \frac{\pop_\unit \cdot \pop_{\altUnit}}{\dist_{\unit\altUnit}},
\]
where $\gravity_u$ is the {\it gravitational constant}, $\dist_{\unit\altUnit}$ is the distance between town $\unit$ and town $\altUnit$, $\pop_\unit$ is the average of $P_{\unit}(t)$ across time, $\overline{\pop}$ is the average of $\pop_{\unit}$ across towns, and $\overline{\dist}$ is the average of $\dist_{\unit\altUnit}$ across all pairs of towns.
The transition processes $N_{EI,\unit}(t)$, $N_{IR,\unit}(t)$ and $N_{QD,\unit}(t)$, for $Q\in\{S,E,I,R\}$,  are modeled as conditional Poisson process with per-capita rates $\mu_{EI,\unit}$, $\mu_{IR,\unit}$ and $\mu_{QD,\unit}$ respectively, and we fix $\mu_{QD,\unit}=(50 \mbox{ year})^{-1}$.
A fraction $\cohort$ of births enter the susceptible cohort on the school admission day, and hence the birth process $N_{BS,\unit}(t)$ is an inhomogeneous Poisson process with rate
$\mu_{BS,\unit}(t-\birthdelay) \big[
  (1-\cohort) + \cohort \delta(t-t_a) \big]$,
where $\mu_{BS,\unit}(t)$ is specified by interpolated census data, $t_a=\lfloor t \rfloor + 252/365$ is the admission date for the year containing $t$, $\delta$ is a Dirac delta function, and $\birthdelay=4\, \mbox{year}$ is a fixed delay between birth and entry into a high-transmission school community.

To describe the measurement process, let $Z_{\unit\comma\time}=N_{IR\comma\unit}(t_\time)-N_{IR\comma\unit}(t_{\time-1})$ be the number of removed infected individuals in the $n$th reporting interval.
Suppose that cases are quarantined once they are identified, so that reported cases comprise a fraction $\rho$ of these removal events.
The case report $\data{y}_{\unit\comma\time}$ is modeled as a realization of a discretized conditionally Gaussian random variable $Y_{\unit\comma\time}$, defined for $y>0$ via
\begin{eqnarray}
\nonumber
\prob\big[Y_{\unit\comma\time}{=}y\mid Z_{\unit\comma\time}{=}z\big] &=& \varPhi\big(y+0.5; \rho_{\unit} z,\rho_{\unit}(1-\rho_{\unit})z+\measOD_{\unit}^2\rho_{\unit}^2z^2\big)
\\
&&\hspace{-20mm}
- \varPhi\big(y-0.5; \rho_{\unit} z,\rho_{\unit}(1-\rho_{\unit})z+\measOD_{\unit}^2\rho^2z^2\big)
\label{eq:obs}
\end{eqnarray}
where $\varPhi(\cdot;\mu,\sigma^2)$ is the $\normal(\mu,\sigma^2)$ cumulative distribution function, and $\measOD$ models overdispersion relative to the binomial distribution.
For $y=0$, we replace $y-0.5$ by $-\infty$ in \eqref{eq:obs}.
Three data points were treated as missing by \citet{he10} due to presumed recording errors.
We followed that decision in the measurement model by including a special value $\mathrm{NA}$ in $\Yspace$ and setting $Y_{\unit,n}$ to be $\mathrm{NA}$ with probability one when $\data{y}_{\unit,n}$ is missing.

We have written the model with all parameters unit-specific, so $\theta=\psi_{1:U}$ with
\arxiv{\begin{equation*}}
\statSinica{$}
\psi_u=\big(\meanBeta_{\unit},\mu_{EI,\unit}, \mu_{IR,\unit},\rho_{\unit},\measOD_{\unit},\sigma_{SE,\unit},\gravity_{\unit},\iota_{\unit},\amplitude_{\unit},\alpha_{\unit}, S_{0,\unit}, E_{0,\unit},I_{0,\unit}\big)
\arxiv{\end{equation*}}
\statSinica{$}
for $u\in\seq{1}{U}$.
This defines an extended model for implementing IBPF with shared parameters, as in \eqref{eq:extension}.
It also lets us address a key data analysis question of which parameters should be shared between units and which should be unit-specific.

\begin{knitrout}
\definecolor{shadecolor}{rgb}{1, 1, 1}\color{fgcolor}\begin{figure}

\includegraphics[width=5.5in]{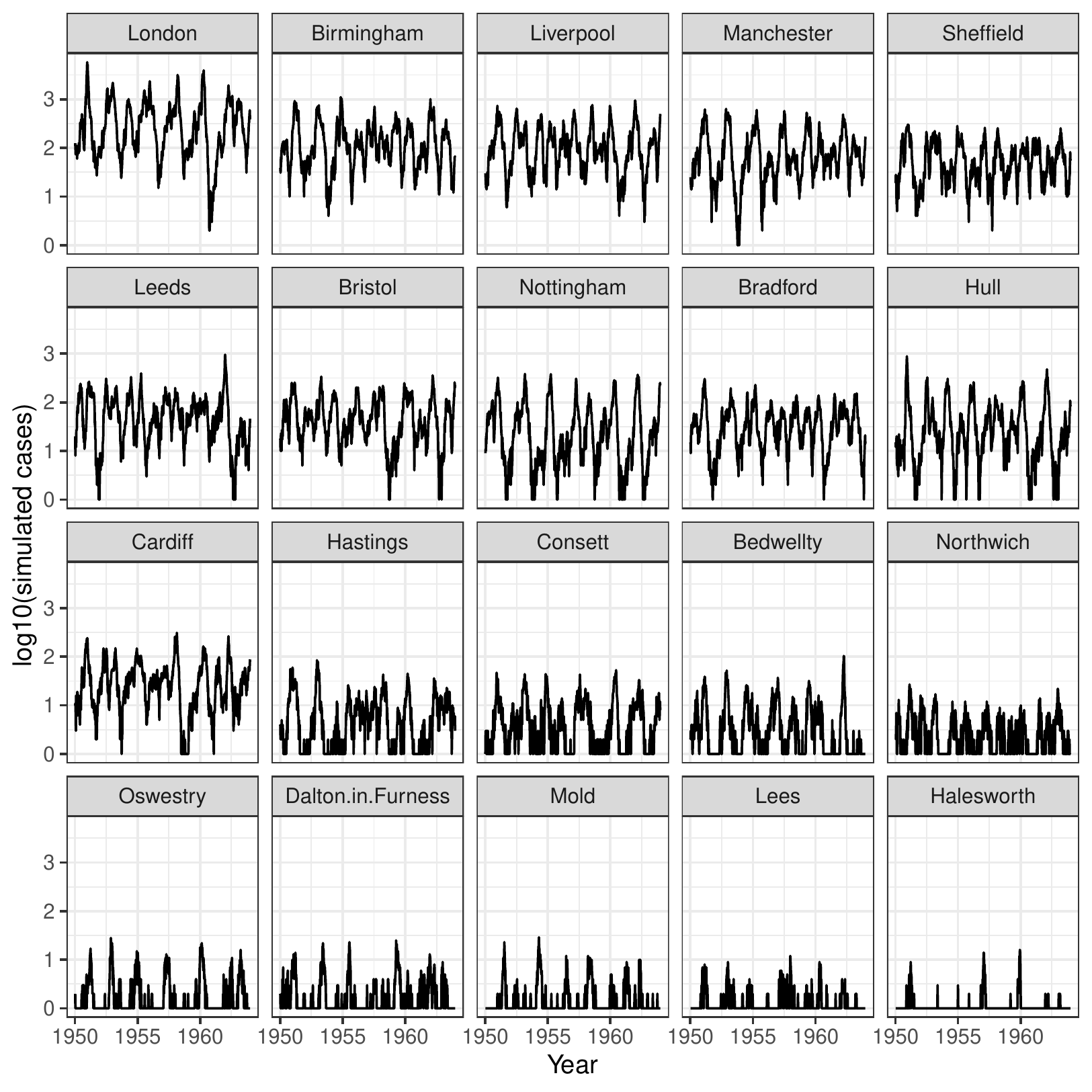} \hfill{}

\caption[Simulated weekly measles case reports]{Simulated weekly measles case reports.}\label{fig:sim-plot}
\end{figure}

\end{knitrout}

The data are shown in Fig~\ref{fig:data-plot} and simulations from the model are shown in Fig~\ref{fig:sim-plot}.
Parameters for the simulated model were based on the analysis of individual towns by \citet{he10}.
In order to investigate estimation of either shared or unit-specific parameters, the simulation was conducted with all parameters being shared.
To find a shared parameter vector capable of providing a reasonable representation of all the towns simultaneously, we experimented to look for a visual match between the data and the simulations.
The parameters used are reported in the supplement (Sec.~\supSecMeaslesParams).

Estimated parameter values may have scientific interest, but we focus on the statistical task of likelihood maximization.
In the presence of weak identifiability, small differences in likelihood could lead to large differences in parameter estimates.
In such situations, a scientist may choose to investigate what functions of the parameters can be inferred accurately without adding additional assumptions, or alternatively they may choose to investigate the consequences of placing constraints on some parameters to improve identifiability of the remainder.
Likelihood maximization permits such investigations, but they are beyond the scope of this article.
Parameter values used for the simulation study and parameter values obtained by the likelihood maximization searches are provided in the supplementary code repository, for the interested reader.

To facilitate data analysis, we wish to have methodology that operates across the full spectrum of decisions on shared versus unit-specific parameter designations.
We therefore test our method on three submodels: {\modelA} has mostly shared parameters, choosing only the initial values and reporting rate to be unit-specific; {\modelB} has every parameter unit-specific; {\modelC} has every parameter is unit-specific and the dynamic coupling (in our context, movement of infected individuals between towns) is replaced by an external forcing of each unit (in our context, an immigration rate of infected individuals from outside the study population).
Sec.~{\supSecMeaslesParams} defines the submodels in more detail.
Submodel {\modelC} provides a useful point of reference since it is a special case of a PanelPOMP model \citep{breto19} and also can be analyzed as a collection of separate POMP models.
We set up our model so that {\modelC} matches the analysis of \citet{he10}.
One may expect that methods which take advantage of the special structure of {\modelC} should out-perform more general methods that permit coupling between units, so we expect the SpatPOMP methods to be less numerically efficient than applying POMP methods separately to each unit.
Questions at stake are: How much less efficient are the SpatPOMP methods? Are simulation-based SpatPOMP inference methods practical for situations such as the measles model of \citet{he10}?

The simulated model is drawn from model {\modelA}, which is nested within model {\modelB} but not within {\modelC}.
For models {\modelB} and {\modelC}, with all parameters are unit-specific, we must estimate $20\times 13=260$ parameters.
This greatly exceeds the 7 shared parameters fitted by \citet{ionides21} for a measles metapopulation model using a large computational effort.
An iterated guided intermediate resampling filter (IGIRF) algorithm was used to fit 9 shared parameters and 3 unit-specific initial value parameters in a measles metapopulation model \citep{park20}.
That analysis used a customized treatment for the unit-specific initial value parameters and did not attempt to estimate other unit-specific parameters.
IGIRF is sensitive to the choice of the guide function, and the model-specific implementation used by \citet{park20} performed better than the generic implementation of IGIRF which is currently available in \code{spatPomp}.
To our knowledge, {\ibpf} is therefore advancing the current limits for scalability of simulation-based maximum-likelihood inference for metapopulation dynamics.

Before engaging in likelihood maximization, one first wants to validate likelihood evaluation.
Likelihood evaluation for metapopulation models was the main topic of \citet{ionides21} and we do not repeat that analysis here.
In brief, the basic particle filter provides a consistent evaluation of the log-likelihood in a limit where the number of particles is large enough to make the Monte Carlo standard deviation small.
This is practical only when $U$ is small (say, $U\le 5$) but that can be used to calibrate the bias induced by BPF, which turns out to be small for our multi-town measles model when each town is its own block.
It may be surprising that resampling independently on each block (which is what the block particle filter does) is able to capture the dependence.
Heuristically, we note that the dynamic dependence between blocks is maintained by BPF, which updates particles according to the full coupled dynamics.
Whether this is sufficient to obtain a good approximation to the filter distribution is situation-dependent, but for the specific case of metapopulation models (for which the strongest coupling is within units rather than between units) the approximation can be empirically successful.
In principle, the approximation error can be reduced by increasing the number of units in each block, but in practice the additional Monte Carlo variance acquired by doing this is not worthwhile in situations where the coupling is relatively weak.
The BPF log-likelihood evaluation at the true parameter value is $\loglik_{\mathrm{true}}=-40612.5$ with a Monte Carlo standard error of $0.6$.

\subsection{IBPF on simulated data}
\label{sec:sim}

\begin{knitrout}
\definecolor{shadecolor}{rgb}{1, 1, 1}\color{fgcolor}\begin{figure}

\includegraphics[width=5in]{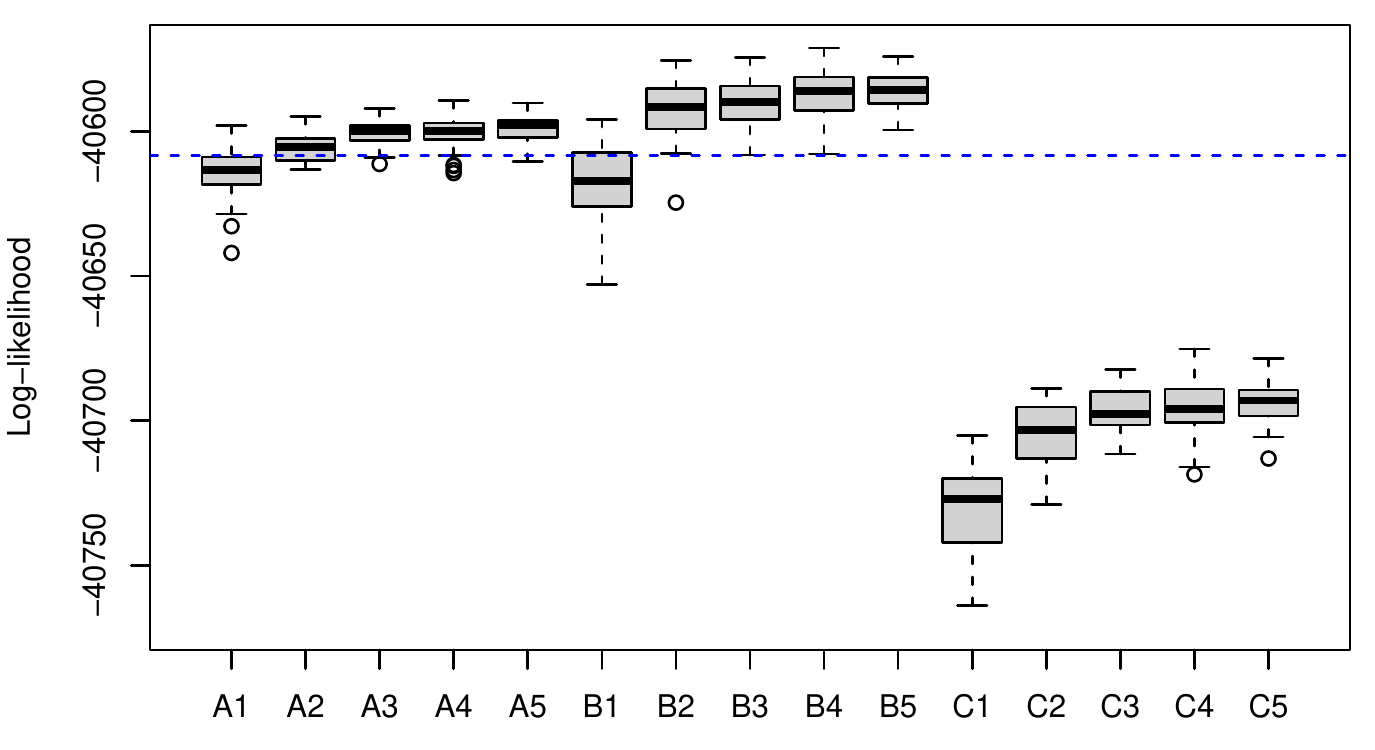} \hfill{}

\caption[Fitting simulated measles data, using an initial search and four refinement steps]{Fitting simulated measles data, using an initial search and four refinement steps. (A1--A5) model \modelA, $4\times 20$ unit-specific parameters and $9$ shared parameters; (B1--B5) model \modelB, $13\times 20$ unit-specific parameters and no shared parameters;  (C1--C5) model \modelC, also all unit-specific, but with immigration rather than coupling, matching He et al (2010). The horizontal dashed line is the log-likelihood at the true parameters, evaluated by BPF.}\label{fig:sim_search_boxplot}
\end{figure}

\end{knitrout}

A goal of this simulation study is to obtain appropriate algorithmic choices for the data analysis.
We seek to develop methodology which is demonstrably successful when the truth is known, before applying it to data.
One could revisit the simulation study based on the data analysis in Sec.~\ref{sec:data} using maximum likelihood estimates of the parameters.

In Fig.~\ref{fig:sim_search_boxplot} we investigate a sequence of successive searches for the MLE for the models {\modelA}, {\modelB} and {\modelC} described above.
Each search was replicated $36$ times.
Search~1 was started with each parameter adjusted by a uniform $[-0.1,0.1]$ random perturbation on an appropriate dimensionless scale (log for non-negative parameters, logit for $[0,1]$ valued parameters, see Sec.~{\supSecAlg}).
This is a fairly small perturbation, but nevertheless sufficient to knock the likelihood of the parameter vector around 200 log units below the MLE (shown in Fig.~\ref{fig:search_diagnostics}).
The goal of this study is to show that the algorithm can succeed reliably on a relatively easy local optimization task.
Subsequent searches were started with four copies of each parameter vector with estimated likelihood in the top 25\% for the previous search.

The MLE is not known exactly in this case.
Wilks' theorem gives an asymptotic expectation that the log-likelihood at the MLE should be greater than the likelihood at the truth by approximately $1/2$ the number of parameters, which here is $(4\times 20+9)/2=44.5$ for {\modelA} and $(13\times 20)/2=130$ for {\modelB}.
The $y$-axis values in Fig.~\ref{fig:sim_search_boxplot} show log-likelihoods exceeding the truth by less than this, indicating some imperfection to the Monte Carlo maximization so far as Wilks' asymptotic result holds in this situation.
Despite this limitation, searches that exceed the likelihood at the truth have found inferentially plausible sets of parameters which can be used to study the likelihood surface around the maximum.
For example, Monte Carlo profile confidence intervals can give proper coverage even in the presence of considerable Monte Carlo error \citep{ionides17,ning21}.

\begin{knitrout}
\definecolor{shadecolor}{rgb}{1, 1, 1}\color{fgcolor}\begin{figure}

\includegraphics[width=5in]{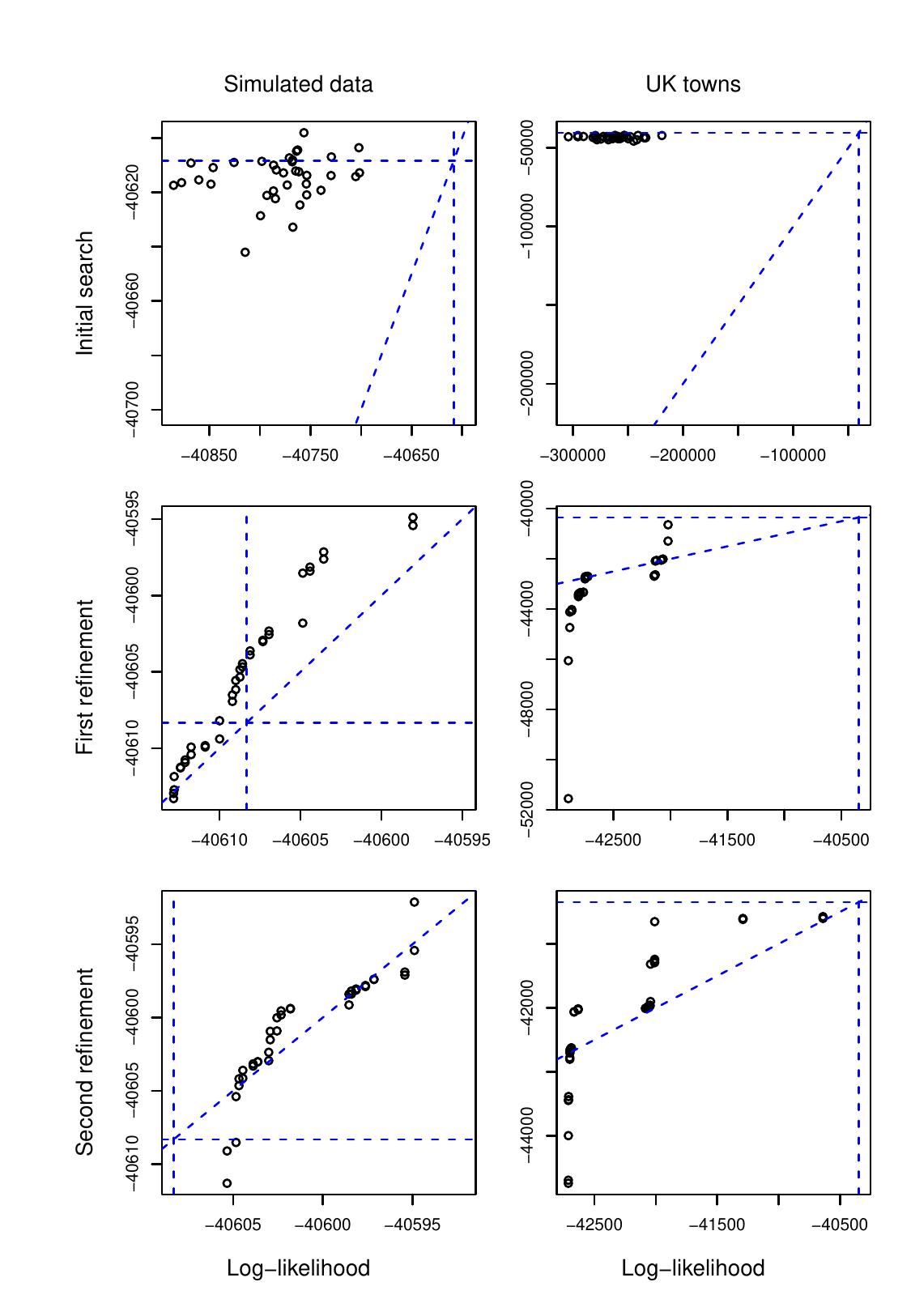} \hfill{}

\caption[Three steps of a likelihood search on simulated data (left panel, model \modelA) and UK measles data (right panel, model \modelB).Dashed lines parallel to the axes show the log-likelihood at the truth (left panel) and the \citet{he10} value (right panel)]{Three steps of a likelihood search on simulated data (left panel, model \modelA) and UK measles data (right panel, model \modelB).Dashed lines parallel to the axes show the log-likelihood at the truth (left panel) and the \citet{he10} value (right panel). Points above the diagonal dashed line show improvement due to the search step.}\label{fig:search_diagnostics}
\end{figure}

\end{knitrout}

In Fig.~\ref{fig:search_diagnostics} we take a closer look at convergence diagnostics corresponding to {\modelA} for the simulated data (the first column) and {\modelB} for the UK measles data (the second column).
For now, we focus on the first column.
The first row plots the log-likelihood obtained after the initial search against the log-likelihood of the randomly selected starting value.
The horizontal and vertical dashed lines denote the likelihood at the truth, and the diagonal dashed line represents equality, so that points above the diagonal show improvement after the search.
We find that {\ibpf} robustly and rapidly approaches a neighborhood of the MLE, as measured by likelihood.
The second row shows that further investigation of the more successful searches can reliably obtain likelihood values higher than those at the truth.
However, for a Monte Carlo search based on Monte Carlo likelihood evaluation, it may be anticipated that pinpointing the exact maximum in a high-dimensional space is problematic.
The third row shows that continued searching does not lead to substantially better outcomes.
When using these methods, we emphasize the need to make proper inferences despite imperfect maximization \citep{ionides17,ning21}.

The results in Fig.~\ref{fig:sim_search_boxplot} and the first column of  Fig.~\ref{fig:search_diagnostics} document that {\ibpf} can be effective on simulated data, but do not guarantee comparable performance for our data analysis.
Indeed, model misspecification that is inevitable for data analysis may be expected to add difficulties to filtering and therefore to numerical methods based on filtering.
Rather, we view the simulation study as a lower bound on the effort required to carry out effective inference on data, and a starting point for investigating effective ways to proceed with the data analysis. Before moving on to the data analysis, we briefly describe the details of the model.

\subsection{IBPF applied to data}
\label{sec:data}

\begin{knitrout}
\definecolor{shadecolor}{rgb}{1, 1, 1}\color{fgcolor}\begin{figure}

\includegraphics[width=5in]{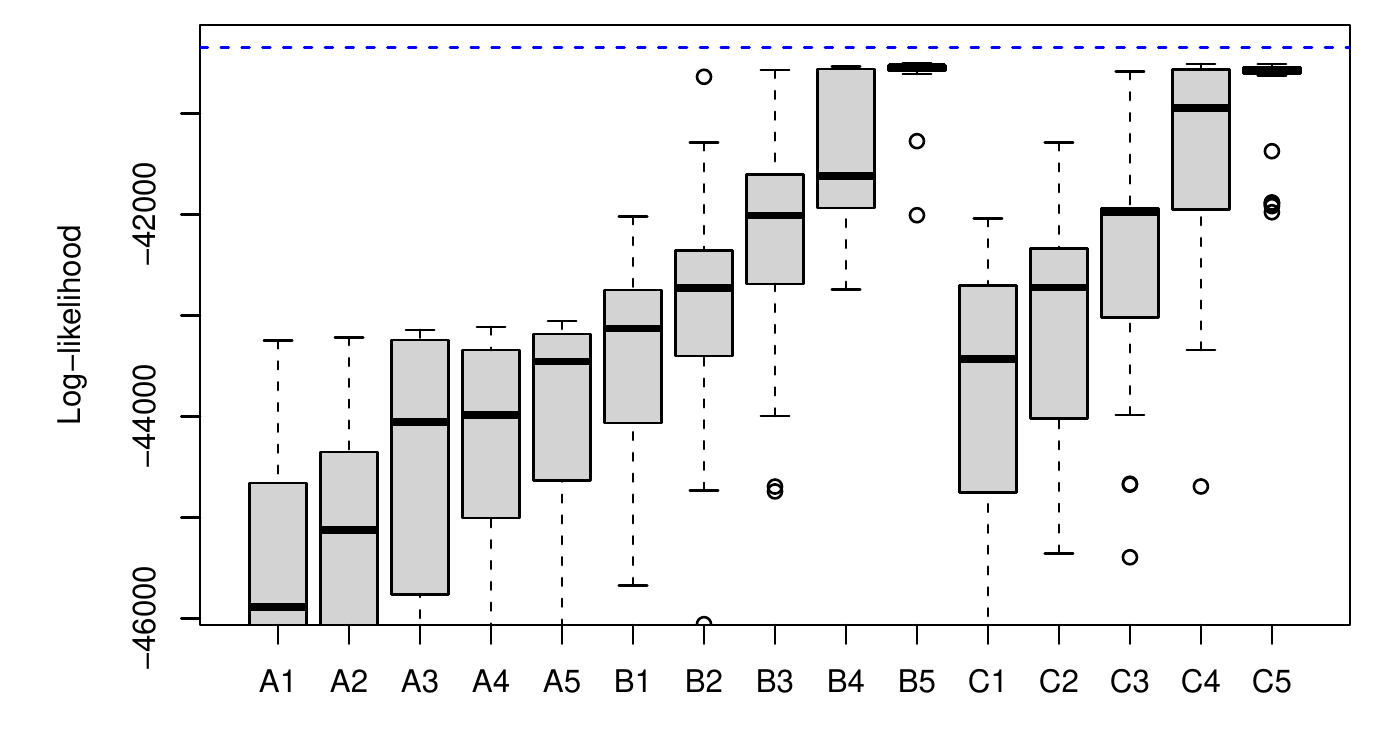} \hfill{}

\caption[Fitting different models to the UK measles data using the method tested on simulated data]{Fitting different models to the UK measles data using the method tested on simulated data. (A) $4\times 20$ unit-specific parameters and $9$ shared parameters; (B) $13\times 20$ unit-specific parameters and no shared parameters. (C) Also all unit-specific, but with immigration rather than coupling. The horizontal dashed line is the likelihood from \citet{he10}.}\label{fig:data_search_boxplot}
\end{figure}

\end{knitrout}

Fig.~\ref{fig:data_search_boxplot} shows results for fitting models {\modelA}, {\modelB} and {\modelC} to UK measles data using successive rounds of {\ibpf}, applying the same method used for the simulation study shown in Fig.~\ref{fig:sim_search_boxplot}.
The dashed lines show the sum of the log-likelihoods obtained by \citet{he10}, $\ell_{\mathrm He}=-40345.7$.
The value $\ell_{\mathrm He}$ corresponds to model {\modelC}.
BPF with each town forming a separate block, applied to an uncoupled model such as {\modelC}, is equivalent to carrying out independent particle filters for each unit.
Indeed, if the published parameter values from \citet{he10} are inserted into the \code{spatPomp} object for model {\modelC}, then the \code{bpfilter} function can be used to carry out BPF, which retrieves $\ell_{\mathrm He}$, up to Monte Carlo error.
Each of models {\modelA}, {\modelB} and {\modelC} in Fig.~\ref{fig:data_search_boxplot} shows shortfall relative to $\ell_{\mathrm He}$.

The largest shortfall is for {\modelA}, and an explanation is that {\modelA} has more shared parameters than the evidence in the data supports.
The results of \citet{he10} suggest that the data are explained better when various parameters are a function of the town population.
However, it remains an open problem to determine suitable functional forms for this relationship, and to establish regularities across towns that can be represented by shared parameters.
This may be investigated using panel methods such as PanelPOMP models \citep{breto19} in addition to consideration of SpatPOMP models.

Models {\modelB} and {\modelC} give rise to comparable likelihoods, which is in contrast to Fig.~\ref{fig:sim_search_boxplot}.
Fitting to simulated data from {\modelB} (in the special case where all unit-specific parameters are equal across all units) the shortfall for {\modelC} in Fig.~\ref{fig:sim_search_boxplot} indicates that we obtain a substantially worse fit by approximating coupling by an uncoupled reservoir of infection.
If the actual data were also explained substantially better by the coupled model, we would expect to see comparable results in the data analysis.
Since we do not, we conclude that this coupled model is not a substantially better explanation than the uncoupled model.
Various other candidate coupling mechanisms have been proposed \citep{bjornstad19} but they have not yet been fitted to the full data, suggesting a lack of appropriate methodology to do so.
Instead, \citet{bjornstad19} looked at summary statistics based on local fade-outs and re-introductions.

We notice from Fig.~\ref{fig:data_search_boxplot} that model {\modelB} has a small but distinct shortfall compared to $\ell_{\mathrm He}$.
On the simulated data, we do not observe this shortfall, and we deduce that the real data provides a more challenging optimization environment.
When carrying out hard optimization problems, it may be possible to develop helpful strategies specific to the model and data in question.
For example, one could try merging unit-specific parameters from different searches, using the likelihood component for each unit to assess successful choices.
Optimization heuristics such as this do not have general theoretical support; if they obtain higher likelihoods that is sufficient justification.
For analysis of panel time series data using PanelPOMP models, such methods have clearer theoretical support and have been found useful \citep{breto19}.

\section{Discussion}
\label{sec:discussion}

Theoretical interest in BPF was inspired by \citet{rebeschini15}, who showed mathematically that BPF can enjoy linear scaling properties under suitable conditions.
\citet{rebeschini15} expressed pessimism about the practical applicability of their algorithm, which may help to explain why the practical use of this algorithm has been limited.
The algorithm was independently invented under the name of a factored particle filter by \citet{ng02} who offered an empirical justification.
However, the proposal of \citet{ng02} also saw only limited use.
The numerical results in Fig.~3 of \citet{ionides21} suggest that BPF is particularly well suited to metapopulation models.
In a metapopulation model, one expects most of the population dynamics to occur at a local level, among the individuals at one spatial unit.
As a broad generalization, in ecological systems dispersion of individuals between spatial units is rare but dynamically important.
Edge effects between blocks, which may be a serious problem for BPF in a system with stronger spatial coupling \citep{ionides21} can therefore be a relatively minor concern for metapopulation models.
Blocks of size one unit are therefore a natural choice for block filtering of metapopulation models when the constituent populations are spatially distinct.

This article has investigated a likelihood maximization approach to inference.
Much research has been done on inference for high-dimensional partially observed stochastic dynamic systems, and we have cited above only the most directly relevant work.
A Bayesian inference approach based on the ensemble Kalman filter was developed by \citet{katzfuss19}.
An expectation-maximization approach based on a block particle filter was presented by \citet{finke17}.
Spatiotemporal models having a convenient factorization across units were studied by \citet{beskos17} and \citet{xu19}.

  We have demonstrated a workflow that led to a likelihood-based assessment of measles metapopulation models ({\modelA} and {\modelB}) with the possibility of finding evidence that they out-perform the uncoupled model, {\modelC}.
  Our methodology showed the potential to successfully refute {\modelC} on simulated data, when the truth was within {\modelA} and {\modelB}, but no advantage was found for these coupled models when the same comparison was carried out on the data.
Likelihood maximization for the measles data fits the {\it common task framework} described by \citet[][Sec.~6]{donoho17}, with the likelihood value for {\modelC} obtained by \citet{he10} providing a benchmark challenge.
Future improvements in models, perhaps facilitated by the open-source models and methods accompanying this article, may obtain metapopulation models that convincingly beat model {\modelC}.

Computation time for Fig.~\ref{fig:sim_search_boxplot} and Fig.~\ref{fig:data_search_boxplot} was approximately 24hr on a single core of a compute node for 4000 particles iterated 100 times over the 730 time points (weekly data for 14 years) for 20 cities.
Each box in these figures involves 36 replications and thus took 24hr on all cores of one node on a computing cluster.
A measure of computational efficiency is the size of problem that can be solved on this timescale, discussed further in Sec.~{\supSecEfficiency}.
Tasks that are considerably larger, perhaps $10^3$ or more spatial units, may require additional approximations such as those inherent in the ensemble Kalman filter \citep{evensen09book,katzfuss19} or other numerical filtering techniques \citep{whitehouse22}.
However, we anticipate that many practical metapopulation analyses can be addressed within the scope we have demonstrated.
The majority of the computational effort is spent on simulating from the dynamic model.
In the \code{spatPomp} R package, as in \code{pomp}, the user supplies a snippet of C code for simulating a single particle between a single arbitrary pair of times, and the package provides a vectorized form of this computation which is used by the inference methodologies within the package.
As a consequence, competitive computational performance is obtained even though the majority of the package is written in R \citep[Table~1]{FitzJohn20}.

\vspace{3mm}


\noindent {\bf Acknowledgments}

\vspace{1mm}

\noindent This work was supported by National Science Foundation grants DMS-1761603 and DMS 1646108, and a Texas A\&M University Seed Fund Grant.
We acknowledge an anonymous referee from a previous article who recommended implementing a block particle filter.
We are grateful to two anonymous referees whose constructive feedback has improved this article.

\arxiv{
\vspace{2mm}
\noindent{\bf Version notes}

\vspace{1mm}

\noindent Version [v1] was submitted to {\it Statistica Sinica}, and [v2] was accepted for publication. This version, [v3], corrects for an omission of the cohort effect in the description of the model in Section~\ref{sec:model}.
}

\InputIfFileExists{ms.bbl}

\arxiv{
\newpage

\setcounter{section}{0}
\setcounter{equation}{0}
\def\theequation{S\arabic{section}.\arabic{equation}}
\renewcommand\thefigure{S\arabic{figure}}
\renewcommand\thetable{S\arabic{table}}
\def\thesection{S\arabic{section}}

\section{Parameters for the measles model}

\newcommand\measlesShared{shared}
\newcommand\measlesUnitSpecific{unit-specific}

\begin{table}[h]
\begin{center}
\begin{tabular}{|l|c|c|c|c|}
\hline
Parameter & \modelA & \modelB & \modelC & Simulation \\
\hline
$\pinit_{S,u}$ & \measlesUnitSpecific & \measlesUnitSpecific & \measlesUnitSpecific &
  $0.032$ \\
$\pinit_{E,u}$ & \measlesUnitSpecific & \measlesUnitSpecific & \measlesUnitSpecific &
  $0.00005$ \\
$\pinit_{I,u}$ & \measlesUnitSpecific & \measlesUnitSpecific & \measlesUnitSpecific &
  $0.00004$ \\
$\rho_u$ & \measlesUnitSpecific & \measlesUnitSpecific & \measlesUnitSpecific &
  $0.5$ \\
$\measOD_{\unit}$ & \measlesShared & \measlesUnitSpecific & \measlesUnitSpecific &
  $0.15$ \\  
$\mu_{EI,\unit}$ & \measlesShared & \measlesUnitSpecific & \measlesUnitSpecific &
  $1 \, \mathrm{week}^{-1}$ \\
$\mu_{IR}$ & \measlesShared & \measlesUnitSpecific & \measlesUnitSpecific &
  $1 \, \mathrm{week}^{-1}$ \\
$\meanBeta_\unit$ & \measlesShared & \measlesUnitSpecific & \measlesUnitSpecific &
  $30 \, \mathrm{week}^{-1}$ \\
$\sigma_{SE,\unit}$ & \measlesShared & \measlesUnitSpecific & \measlesUnitSpecific &
  $0.15 \, \mathrm{year}^{1/2}$ \\
$\amplitude_{\unit}$ & \measlesShared & \measlesUnitSpecific & \measlesUnitSpecific &
  $0.5$ \\
$\alpha_{\unit}$ & \measlesShared & \measlesUnitSpecific & \measlesUnitSpecific &
  $1$ \\
$\cohort$& \measlesShared & \measlesUnitSpecific & \measlesUnitSpecific &
  $0$ \\  
$\gravity_{\unit}$ & \measlesShared & \measlesUnitSpecific & 0 &
  $400$ \\
$\iota_{\unit}$ & 0 & 0 & \measlesUnitSpecific &
  0 \\
\hline
\end{tabular}
\caption{Parameters for the measles model. Sub-model {\modelA} has 4 unit-specific parameters and 9 shared parameters, with movement between units following a gravity equation. Sub-model {\modelB} has 13 unit-specific parameters and no shared parameters, with gravity movement. Sub-model {\modelC} has 13 unit-specific parameters and no shared parameters, with independent immigration of infections rather than movement between units. The last column shows the parameter values used for the simulated data in Figure~2 of the main text.
}\label{tab:params}
\end{center}
\end{table}

\noindent Notes on Table~\ref{tab:params}:
\begin{enumerate}
\item In the model formulation, all parameters are written as unit-specific, i.e., with a $u$ subscript. Shared parameters take equal values across all units.
\item For the simulation, all parameters are shared, so one can assess the inferential consequences of estimating models with some or all parameters unit-specific. Establishing a good choice of parameters to be shared or unit-specific for the data is a data analysis goal.
\item The initial values of the latent states are parameterized as fractions of the total population, so
$S_{0,u}=\pinit_{S,u}P_u(t_0)$,
$E_{0,u}=\pinit_{E,u}P_u(t_0)$,
$I_{0,u}=\pinit_{I,u}P_u(t_0)$,
and $R_{0,u}=\big(1-\pinit_{S,u}-\pinit_{E,u}-\pinit_{I,u}\big)P_u(t_0)$.
\item The parameterization used in the numerical implementation replaces $\meanBeta_{\unit}$ with the basic reproductive number, a dimensionless ratio defined as
$\mathscr{R}_{0,\unit}= \meanBeta_{\unit} \big(\mu_{IR,\unit}+\mu_D\big)^{-1}$.
For the simulation,  $\mathscr{R}_{0,\unit}=30$.
\end{enumerate}


\section{Algorithmic parameters and transformations}


\begin{table}[h]
\begin{center}
\begin{tabular}{|l|c|c|c|c|}
\hline
Parameter & Value \\
\hline
$J$ & 4000 \\
$M$ & 100 \\
$\breve\sigma$ & 0.005 (0.00125 on some searches for the simulated data)\\
$\spatReg$ & 0.1 \\
$a$ & 0.5 \\
\hline
\end{tabular}
\caption{Algorithmic parameters used for applying IBPF to the measles model. The same values were used for each of the sub-models {\modelA}, {\modelB} and {\modelC}.}\label{tab:algpars}
\end{center}
\end{table}

\noindent Notes on Table~\ref{tab:algpars}:
\begin{enumerate}
\item The full $D\times \Time$ matrix of perturbation standard deviations with entries $\sigma_{d,n}$ was reduced to a single algorithmic parameter, $\breve\sigma$, via
\begin{equation}
\nonumber
\sigma_{d,n} =
  \left\{ \begin{array}{ll}
    \breve\sigma & \mbox{if $\theta_d$ is not an IVP and not $\alpha_{\unit}$}
    \\
    0.1\, \breve\sigma & \mbox{if $\theta_d$ is $\alpha_{\unit}$}
    \\
    2\, \breve\sigma & \mbox{if $\theta_d$ is an IVP, and $n=0$}
    \\
    0 & \mbox{if $\theta_d$ is an IVP, and $n\ge 1$}
  \end{array}\right.
\end{equation}
where the initial value parameters (IVPs) are $\pinit_{S,u}$, $\pinit_{E,u}$ and $\pinit_{I,u}$.

\item Parameters perturbations were carried out on a transformed scale. A logit transform was used for $\pinit_{S,u}$, $\pinit_{E,u}$, $\pinit_{I,u}$, $\rho_{\unit}$. A log transformation was used for $\measOD_{\unit}$, $\sigma_{SE,u}$, $\mu_{EI,u}$, $\mu_{IR,\unit}$, $\meanBeta_{\unit}$, $\alpha_{\unit}$, $\gravity_{\unit}$, $\iota_{\unit}$. No transformation was used for $\amplitude_{\unit}$.

\item The algorithmic parameters do not have any scientific significance once successful maximization has been demonstrated. They may affect the ease of successful maximization, or even the ability to attain this within an acceptable level of Monte Carlo uncertainty.

\item $M=100$ was chosen empirically.
For a computationally challenging maximization problem, we expect to carry out many searches from a variety of starting values, and we conduct further experiments following up on successful leads.
In this case, it is enough to choose $M$ so that each search has a fair chance of finding a higher likelihood when there are local improvements to be made.

\item Numerical experiments were carried out to choose $J$.
We look for the smallest $J$ such that there is not much to be gained by making $J$ larger.
This is carried out by looking for evidence about the bias and variance (discussed further in Sec.~\supSecEfficiency) which both need consideration when carrying out likelihood ratio tests and Monte Carlo adjusted profile confidence intervals \citep{ionides17,ning21}.

\item Each iteration of IBPF produces a log-likelihood estimate corresponding to the extended model with dynamically perturbed parameters. At the end of the search, the log-likelihood was re-evaluated using BFP, with 10 replications at $J=8000$ particles. A high level of effort on likelihood evaluation assists the task of building understanding about the likelihood surface from repeated Monte Carlo searches.

\item The exact record of all our computations is the source code for our numerical results, which is available at \url{https://github.com/ionides/ibpf_article}.

\end{enumerate}


\section{Computational efficiency}

Numerical error of Monte Carlo methods can be decomposed as bias and variance, and efficiency corresponds to how error scales with the amount of computational resources used. Expended resources can be quantified by objective metrics such as joules or dollars, but in practice we usually assess resources in terms of computational time on the machines that we personally have available.

Here, our main goals are evaluation and maximization of the log-likelihood.
Variance in Monte Carlo likelihood estimates results in negative bias on the log-likelihood due to Jensen's inequality.
Approximations involved in constructing a filter provide another source of bias for estimating the log-likelihood.
This Monte Carlo approximation bias has negative expectation, over when the model is correct, since log-likelihood is a proper scoring rule \citep{gneiting07}.
Monte Carlo maximization error (meaning the difference between the unknown, exact log-likelihood at the Monte Carlo MLE compared to the unknown, exact MLE) can only be negative.
Based on these considerations, we seek methods giving high average Monte Carlo log-likelihood at a Monte Carlo MLE, and our practical goal is to obtain reliably high log-likelihoods using calculations taking no longer than a day or so on one 36-core computing node.


\section{Varying the spatial autoregression parameter, $\spatReg$}

\fontsize{9}{11.5pt plus.8pt minus .6pt}\selectfont

\begin{knitrout}
\definecolor{shadecolor}{rgb}{1, 1, 1}\color{fgcolor}\begin{figure}[h]

\includegraphics[width=4.5in]{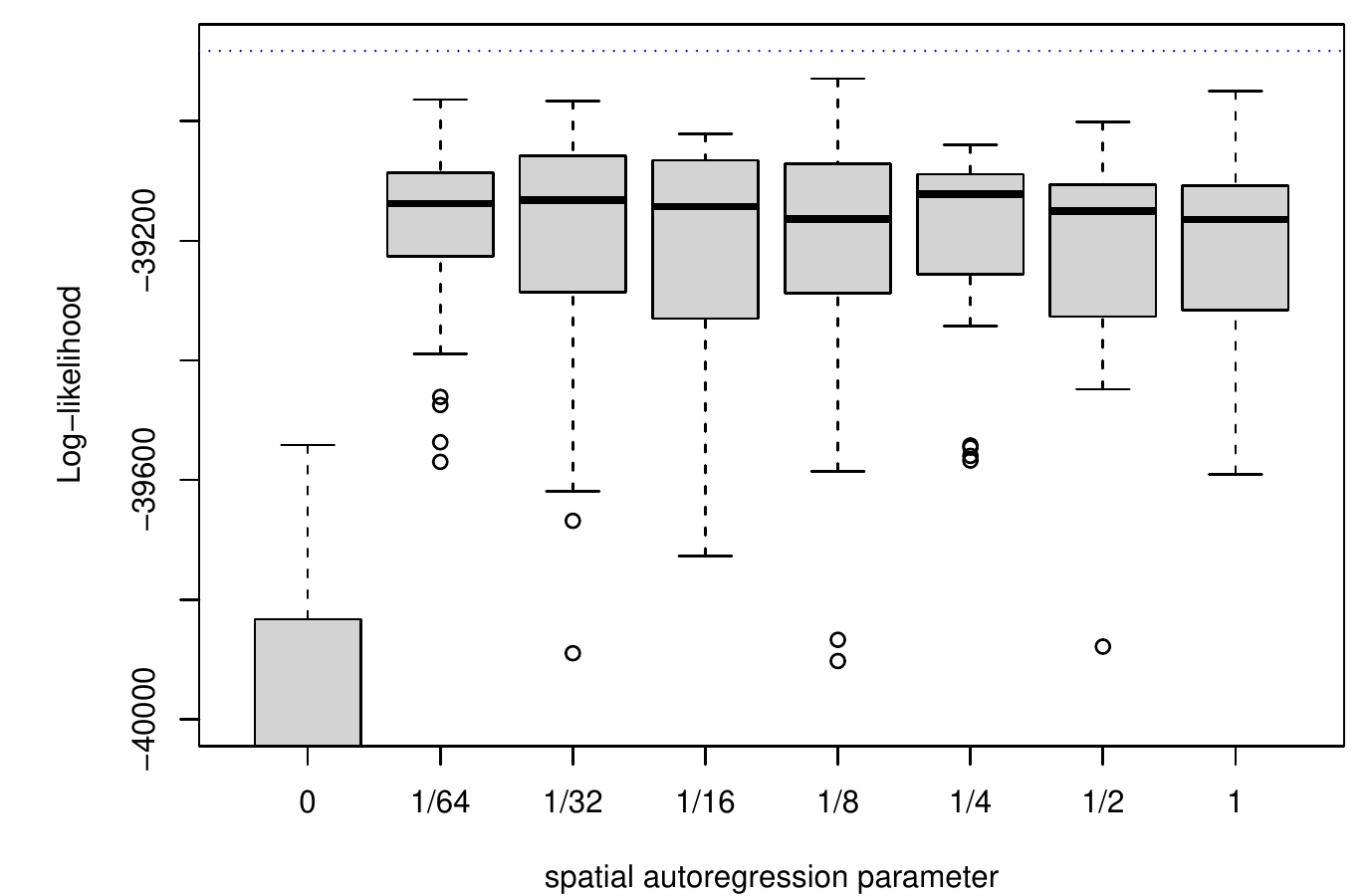} \hfill{}

\caption[Varying the IBPF spatial autoregression algorithmic parameter for shared parameters with simulated measles data]{Varying the IBPF spatial autoregression algorithmic parameter for shared parameters with simulated measles data. The log-likelihood was obtained using $M=100$ iterations of IBPF starting at random parameters (median starting log-likelihood, -275000). The horizontal dashed line denotes the log-likelihood at the true parameters.}\label{fig:spatreg_boxplot}
\end{figure}

\end{knitrout}

}

\end{document}

%% file: header.tex
\newcommand\ibpf{IBPF}

\newcommand\supSecMeaslesParams{S1}
\newcommand\supSecAlg{S2}
\newcommand\supSecEfficiency{S3}
\newcommand\supFigSpatReg{S1}

\usepackage{enumerate,alltt,xstring}
\usepackage[ruled,noline,linesnumbered]{algorithm2e}
\usepackage{setspace}
\SetKwFor{For}{for}{do}{end}
\SetKwFor{While}{while}{do}{end}
\SetKwInput{KwIn}{input}
\SetKwInput{KwOut}{output}
\SetKwInput{KwCplx}{complexity}
\SetKwInput{KwIndices}{note}
\SetKwBlock{Begin}{procedure}{}
\DontPrintSemicolon

\newcommand\amplitude{h}
\newcommand\meanBeta{{\overline\beta}}
\newcommand\cohort{c}

\newcommand\gravity{G}
\newcommand\pop{\mathrm{pop}}  
\newcommand\schoolTermFrac{s}
\newcommand\pinit{\pi}
\newcommand\birthdelay{\Delta}

\newcommand\siOnly[1]{} 

\newcommand\code[1]{\texttt{#1}}
\usepackage{url}

\usepackage{color}
\usepackage[normalem]{ulem}
\definecolor{orange}{rgb}{1,0.5,0}

\definecolor{green}{rgb}{0,0.5,0}

\definecolor{cyan}{rgb}{0,.5,.5}

\newcommand\Xspace{{\mathbb X}}
\newcommand\Yspace{{\mathbb Y}}

\newcommand\data[1]{#1^*}

\newcommand\comma{{\hspace{-0.25mm},\hspace{-0.2mm}}}

\usepackage[mathscr]{euscript}
\newcommand\modelA{$\mathscr{A}$}
\newcommand\modelB{$\mathscr{B}$}
\newcommand\modelC{$\mathscr{C}$}
\newcommand\measOD{\tau}

\newcommand\unit{u}
\newcommand\altUnit{\tilde u}
\newcommand\Unit{U}

\usepackage[mathscr]{euscript}

\renewcommand\time{n}
\newcommand\myvec[1]{\boldsymbol{#1}}

\newcommand\Time{N}
\newcommand\Np{J}
\newcommand\np{j}
\newcommand\altNp{k}

\newcommand\resampleIndex{i}

\newcommand\spatReg{r}

\newcommand\Nit{M}
\newcommand\nit{m}

\newcommand\block{k}
\newcommand\Block{K}
\newcommand\blockweight{w}
\newcommand\blocklist{\mathcal{B}} 

\newcommand\prob{\mathbb{P}}

\newcommand\given{{\,\vert\,}}

\newcommand\mycolon{{\hspace{0.6mm}:\hspace{0.6mm}}}
\newcommand\seq[2]{{#1}\mycolon{#2}}
\newcommand\mydot{{\,\cdot\,}}

\newcommand\giventh{{\hspace{0.5mm};\hspace{0.5mm}}}
\newcommand\normal{\mathcal{N}}

\newcommand\loglik{\lambda}

\newcommand\param{\,;}

\newcommand\dist{d}

\def\loglik{\ell}

\usepackage{amsthm}

